\documentclass[prd,showpacs,showkeys]{revtex4}
\usepackage{graphicx}% Include figure files
\usepackage{epsfig}
\usepackage{dcolumn}% Align table columns on decimal point
\usepackage{bm}% bold math
\usepackage{amssymb}
\usepackage[latin1]{inputenc}

\begin{document}
%\preprint{}
%\tightenlines
\title{Series expansion of the photon self-energy in QED and
the photon anomalous magnetic moment}
\author{ H. P\'erez Rojas$^{\dag}$ and E. Rodriguez Querts$^{\ddag}$ }
\affiliation{Instituto de Cibernetica, Matematica y Fisica, Calle E
309, Vedado, Ciudad Habana, Cuba. \\$^{\dag}$ hugo@icmf.inf.cu,
$^{\ddag}$ elizabeth@icmf.inf.cu}
% \email{hugo@icmf.inf.cu, elizabeth@icmf.inf.cu}
\begin{abstract}
We start from the analytical expression of the eigenvalues
$\kappa^{(i)}$ of the photon self-energy tensor in an external
constant magnetic field $B$ calculated by Batalin  Shabad in the
Furry representation, and in the one-loop approximation. We expand
in power series of the external field and in terms of the squared
photon transverse momentum $z_2$ and (minus) transverse energy
$z_1=k^2-z_2$, in terms of which are expressed $\kappa^{(i)}$. A
general expression is given for the photon anomalous magnetic
moment $\mu_{\gamma}>0$ in the region of transparency, below the
first threshold for pair creation, and it is shown that it is
positive, i.e. paramagnetic. The results of the numerical
calculation for $\mu_{\gamma}>0$ are displayed in a region close
to the threshold.
\end{abstract}


\maketitle
 This paper is to be understood as a complement of the
shorter paper \cite{Selym}. Here calculations of several
quantities appearing in it are made in more detail. We start from
the renormalized eigenvalues of the polarization operator in
presence of a constant homogeneous magnetic field in one-loop
approximation, given by Shabad in \cite{Shabad}:
\section{Polarization operator}
\begin{equation}\label{op-pol}
\kappa_{i}=\frac{e^2}{2\pi ^2}\int_{0}^{\infty}d\tau
\int_{-1}^{1}d\eta e^{-m^2 \tau}\left[\rho_{i}\frac{eB}{\sinh (eB
\tau)} e^{-z_{2}\frac{\sinh (eB[(1+\eta)/2] \tau)\sinh
(eB[(1-\eta)/2] \tau)}{eB \sinh(eB \tau)}-z_1 \frac{1-\eta
^2}{4}\tau}+(z_1 +z_2)\frac{1-\eta ^2}{8\tau} \right],
\end{equation}

where we have
\begin{eqnarray*}
  \rho_1 &=& -\frac{z_1
+z_2}{2}\frac{\sinh (eB[(1+\eta)/2] \tau)\cosh (eB[(1+\eta)/2]
\tau)}{\sinh(eB \tau)}\frac{1-\eta}{2}, \\
  \rho_2 &=& -\frac{z_1}{2} \cosh(eB\tau)\frac{1-\eta ^2}{4}-\frac{z_2}{2}
  \frac{\sinh (eB[(1+\eta)/2] \tau)\cosh (eB[(1+\eta)/2]
\tau)}{\sinh(eB \tau)}\frac{1-\eta}{2}, \\
  \rho_3 &=& -\frac{z_2}{2}\frac{\sinh (eB[(1+\eta)/2] \tau)\sinh (eB[(1-\eta)/2]
\tau)}{\sinh^{2}(eB \tau)} -\frac{z_1}{2}\frac{\sinh (eB[(1+\eta)/2]
\tau)\cosh (eB[(1+\eta)/2] \tau)}{\sinh(eB \tau)}\frac{1-\eta}{2},
\end{eqnarray*}
where  $z_1 =k_{\parallel}^2 -k_{0}^2$ and $ z_2 =k_{\perp}^2$ \\
After introducing the new variable $t=eB\tau$, previous expressions
look like
\begin{equation}\label{op-pol2}
\kappa_{i}=\frac{e^2}{2\pi ^2}\int_{0}^{\infty}dt \int_{-1}^{1}d\eta
e^{-\frac{m^2}{eB}t
 }\left[\frac{\rho_{i}}{\sinh t}
e^{\zeta}+(z_1 +z_2)\frac{1-\eta ^2}{8t} \right],
\end{equation}
\begin{eqnarray*}
\zeta&=&-\frac{z_{2}}{eB}\frac{\sinh ([(1+\eta)/2] t)\sinh
([(1-\eta)/2]
t)}{\sinh t}-\frac{z_1 }{eB}\frac{1-\eta ^2}{4}t\\
  \rho_1&=&-\frac{z_1
+z_2}{2}\frac{\sinh ([(1+\eta)/2] t)\cosh ([(1+\eta)/2] t)}{\sinh
t}\frac{1-\eta}{2}
\\
  \rho_2&=&-\frac{z_1}{2} \cosh t\frac{1-\eta ^2}{4}-\frac{z_2}{2}
  \frac{\sinh ([(1+\eta)/2] t)\cosh ([(1+\eta)/2]
t)}{\sinh t}\frac{1-\eta}{2}
\\
  \rho_3&=&-\frac{z_2}{2}\frac{\sinh ([(1+\eta)/2] t)\sinh ([(1-\eta)/2]
t)}{\sinh^{2} t} -\frac{z_1}{2}\frac{\sinh ([(1+\eta)/2] t)\cosh
([(1+\eta)/2] t)}{\sinh t}\frac{1-\eta}{2}
\end{eqnarray*}
\subsection{Small frequencies limit $z_1 ,z_2<<m^2$}
We are interested in a wide range of frequencies characterized by
the condition $z_1 ,z_2<<m^2$. We can express (\ref{op-pol2}) as
\begin{equation}\label{series-op-pol1}
\kappa_{i}=\frac{e^2}{2\pi ^2} \int_{0}^{\infty}dt
\int_{-1}^{1}d\eta e^{-\frac{m^2}{eB}t } \left[\frac{\rho_{i}}{\sinh
t} \sum_{l=0}^{\infty} \frac{\zeta^{l}}{l!} +(z_1 +z_2)\frac{1-\eta
^2}{8t} \right],
\end{equation}
or
\begin{equation}\label{series-op-pol}
 \kappa_{i}=\sum_{l=0}^{\infty}
\kappa_{i}^{(l)},
 \end{equation}
\begin{eqnarray*}
   \kappa_{i}^{(0)} &=&
\frac{e^2}{2\pi ^2} \int_{0}^{\infty}dt \int_{-1}^{1}d\eta
e^{-\frac{m^2}{eB}t } \left[\frac{\rho_{i}}{\sinh  t} +(z_1
+z_2)\frac{1-\eta ^2}{8t} \right], \\
\kappa_{i}^{(l)} &=& \frac{e^2}{2\pi ^2} \frac{1}{l!}
\int_{0}^{\infty}dt \int_{-1}^{1}d\eta e^{-\frac{m^2}{eB}t}
\frac{\rho_{i}}{\sinh t} \zeta^{l},  \hspace{1cm} l=1,2,3,...,
\end{eqnarray*}
and consider  the first few terms of $\kappa_{i}^{(l)}$ in the
series expansion (\ref{series-op-pol}). Here we will explicitly
compute $\kappa_{i}^{(0)}$ and $\kappa_{i}^{(1)}$.

To get the linear  term  of $\kappa_{i}$ in $z_1, z_2$ (which for
$z_1= z_2$ will be used as the infrared limit), we take $e^{\zeta
=1}$, and from $\rho_{i}$ we have, by naming it as
$\kappa_{i}^{(0)}$,
\begin{eqnarray}
    \kappa_{1}^{(0)} &=& (z_{1}+z_{2}) \frac{e^2}{2\pi ^2}
    \int_{0}^{\infty}dt
 \int_{-1}^{1}d\eta e^{-\frac{m^2}{eB}t }
\left[\frac{1+\eta}{2 t}-\frac{\sinh ([(1+\eta)/2] t)\cosh
([(1+\eta)/2] t)}{\sinh ^{2}
t}\right]  \frac{1-\eta}{4}\\
\nonumber
    \kappa_{2}^{(0)} &=& z_{1} \frac{e^2}{2\pi ^2}
    \int_{0}^{\infty}dt
 \int_{-1}^{1}d\eta e^{-\frac{m^2}{eB}t }
\left[\frac{1}{t} - \frac{\cosh t}{\sinh t}
\right] \frac{1-\eta ^2}{8}+ \\
  & & \hspace{2cm}+z_{2} \frac{e^2}{2\pi ^2}  \int_{0}^{\infty}dt
\int_{-1}^{1}d\eta e^{-\frac{m^2}{eB}t}  \left[\frac{1+\eta}{2
t}-\frac{\sinh ([(1+\eta)/2] t)\cosh ([(1+\eta)/2] t)}{\sinh ^{2}
t}\right]  \frac{1-\eta}{4},
\end{eqnarray}
\begin{eqnarray}\nonumber
 \kappa_{3}^{(0)} &=& z_{1} \frac{e^2}{2\pi ^2} \int_{0}^{\infty}dt
\int_{-1}^{1}d\eta e^{-\frac{m^2}{eB}t } \frac{1-\eta }{4}
\left[\frac{1+\eta}{2 t}
 -\frac{\sinh ([(1+\eta)/2] t)\cosh([(1+\eta)/2] t)}{\sinh ^{2} t}
\right]+ \\
 & & \hspace{2.8cm}+z_{2} \frac{e^2}{2\pi ^2}  \int_{0}^{\infty}dt
\int_{-1}^{1}d\eta e^{-\frac{m^2}{eB}t } \frac{1}{2}
\left[\frac{1-\eta ^{2}}{4 t}-\frac{\sinh ([1+\eta)/2] t)\sinh
([1-\eta)/2] t)}{\sinh ^{3} t}\right],
\end{eqnarray}
After integrating in $\eta$ we get
\begin{equation}\label{k0}
   \kappa_{i}^{(0)} = z_{1} \frac{e^2}{2\pi ^2}
    \int_{0}^{\infty}dt
 e^{-\frac{m^2}{eB}t }
g_{i1}+ z_{2} \frac{e^2}{2\pi ^2}  \int_{0}^{\infty}dt
 e^{-\frac{m^2}{eB}t}
 g_{i2}  \hspace{1cm} i=1,2,3
\end{equation}
\begin{eqnarray*}
     g_{11} &=& g_{12} = g_{22}  = g_{31}  =-\frac{1}{12\,t } - \frac{\coth t }{4\,{t}^2} +
\frac{\coth^2 t }{4\,t }>0, \\
   g_{21} &=&\frac{1}{6t} - \frac{\cosh t}{6\sinh t}<0, \\
   g_{32}  &=& \frac{1}{6 t }+\frac{1}{2 t
\sinh^2 t} -\frac{\cosh t}{2 \sinh^3 t} >0.
\end{eqnarray*}

Now we are interested in the next order quadratic on $z_1, z_2$. Let
us call this term  $\kappa_{i}^{(1)}$. This term in
(\ref{series-op-pol}) is
\begin{eqnarray}
    \kappa_{1}^{(1)} &=&  \frac{e^2}{4\pi ^2 eB}
\int_{0}^{\infty}dt
  e^{-\frac{m^2}{eB}t } \left(z_{1}^2
j_{11} +z_1z_2 j_{12}+z_{2}^2 j_{22} \right), \\
    \kappa_{2}^{(1)} &=&
   \frac{e^2}{4\pi ^2 eB}
\int_{0}^{\infty}dt
  e^{-\frac{m^2}{eB}t} \left(z_{1}^2
h_{11} +z_1z_2 h_{12}+z_{2}^2 h_{22} \right),
 \\
    \kappa_{3}^{(1)} &=&
   \frac{e^2}{4\pi ^2 eB}
\int_{0}^{\infty}dt
  e^{-\frac{m^2}{eB}t } \left(z_{1}^2
f_{11} +z_1z_2 f_{12}+z_{2}^2 f_{22} \right),
  \end{eqnarray}
  where
  \begin{eqnarray*}
    \nonumber j_{11}
&=& \int_{-1}^{1}d\eta \frac{(1-\eta)(1-\eta^2) }{8}t
\frac{\sinh([(1+\eta)/2]t)\cosh([(1+\eta)/2]t)}{\sinh ^{2} t}\\
&=& -\frac{1}{4\,(t )^2} - \frac{3\,\coth t }{4\,(t )^3}
   + \frac{3\,{\coth^2 t}}{4\,(t )^2}, \\
   \nonumber
   j_{12} &=& \int_{-1}^{1}d\eta
\frac{\sinh([(1+\eta)/2]t)\cosh([(1+\eta)/2]t)}{\sinh ^{2}
t}\frac{(1-\eta) }{2} \left( \frac{\sinh ([(1+\eta)/2] t) \sinh
([(1-\eta)/2] t)}{\sinh
t}+\frac{1-\eta^2 }{4}t \right) \\
&=& -\frac{1}{8 \sinh^2 t }+\frac{19}{32 (t )^2 \sinh^2
  t
  }+\frac{11 \cosh 2 \tau'}{32 (\tau' )^2 \sinh^2 t} -\frac{3 \coth  t}{16 t \sinh^2
  t}-\frac{3 \sinh 2 t}{8 (t )^3 \sinh^2 t}
 \\
\nonumber
 j_{22} &=& \int_{-1}^{1}d\eta\frac{\sinh^2([(1+\eta)/2]\tau')\sinh([(1-\eta)/2]t)\cosh([(1+\eta)/2]t)}{\sinh ^{3}
t}\frac{(1-\eta) }{2}
\\
&=& \frac{3 \coth^2 t}{16 (t)^2} -\frac{1}{8 \sinh^2 t}
  -\frac{3 \coth t}{16 t \sinh^2 t}\\
\nonumber
h_{11} &=&  \int_{-1}^{1}d\eta\frac{(1-\eta^2)^2 }{16}t \coth t,\\
&=& \frac{t}{15}\coth t \\
 \nonumber
 h_{12} &=& \int_{-1}^{1}d\eta\frac{1-\eta^{2}
}{4}\frac{\sinh ([(1+\eta)/2]t)}{\sinh ^{2} t}
\left( \sinh ([(1-\eta)/2] t)\cosh t+\cosh([(1+\eta)/2]t)\frac{1-\eta }{2}t \right), \\
&=&\frac{1}{12} \left(2-\frac{3\coth t}{ (t)
  ^{3}}+\left(2+\frac{3}{(t)^2}\right)\frac{1}{\sinh^2 t}\right)
 \\
\nonumber
 h_{22} &=& \int_{-1}^{1}d\eta \frac{1-\eta }{2}\frac{\sinh^2 ([(1+\eta)/2]t)\sinh ([(1-\eta)/2]t)\cosh([(1+\eta)/2]t)}
 {\sinh ^{3} t},\\
 &=& \frac{1}{16 (t)^2} \left(3\coth^2 t-
  t\left(2 t +3\coth t \right)\frac{1}{\sinh^2 t}\right)\\
 \nonumber
 f_{11} &=& \int_{-1}^{1}d\eta \frac{(1-\eta^2)(1-\eta)}{8}\frac{\sinh([(1+\eta)/2]t)\cosh([(1+\eta)/2]t)}
 {\sinh ^{2}t}t,\\
 &=& -\frac{1}{4 (t)^3} \left(t+3\coth t- 3t\coth^2 t\right)\\
\nonumber
 f_{12} &=& \int_{-1}^{1}d\eta \frac{1-\eta }{2}\frac{\sinh
([(1+\eta)/2]t)\sinh([(1-\eta)/2]t)}{\sinh ^{3} t}
\left( \sinh ([(1+\eta)/2] t)\cosh ([(1+\eta)/2] t)+\frac{1+\eta }{2}t \right), \\
&=&\frac{1}{48 (t)^2} \left(9+\left(33-6(t)^2+\tau'(-33+8(t)^2)\coth
t \right)
  \frac{1}{\sinh^2 t}\right)
 \\
\nonumber
 f_{22} &=& \int_{-1}^{1}d\eta\frac{\sinh^2 ([(1+\eta)/2]t)\sinh^2 ([(1-\eta)/2] t)}
 {\sinh ^{4} t}\\
 &=& \frac{1}{8\sinh^4(t)} \left(4+2\cosh(2 t)-
  \frac{3}{t}\sinh (2t) \right)
  \end{eqnarray*}

\subsection{Small fields limit $eB<<m^2$}
For fields $eB<<m^2$ (actually it is enough that $\frac{eB}{m^2}
\lesssim 10^{-1}$) the functions inside the integrals in
$\kappa_{i}^{(0),(1)}$ are significantly different from zero only
for $t<<1$ and we can expand the following expressions  around $t=0$
and retain the first few terms:

\begin{eqnarray} \label{g}
g_{11}&\approx& \frac{t }{45} - \frac{{t }^3}{315} +
\frac{2\,{t}^5}{4725} - \frac{{t }^7}{18711}
,\\
  g_{21} &\approx& \frac{1}{6 } \left(-\frac{t}{3}+\frac{t^3}{45}
  -\frac{2 t^5}{945}+\frac{t^7}{4725}\right),\\
    g_{32}&\approx&
\frac{t}{15}-\frac{t^3}{63}
  +\frac{2 t^5}{675}-\frac{t^7}{2079},
\end{eqnarray}
and
\begin{eqnarray}
  j_{11}
   &\approx&\frac{1}{15} - \frac{t ^2}{105} + \frac{2\,t^4}{1575} - \frac{t^6}{6237} +
  \frac{1382\,t^8}{70945875} ,\\
   j_{12} &\approx& \frac{2}{15} - \frac{t^2}{42} + \frac{t^4}{270} - \frac{82\,t^6}{155925} +
  \frac{907\,t ^8}{12899250} ,
  \\
  j_{22}
  &\approx& \frac{1}{15} - \frac{t ^2}{70} + \frac{23\,t ^4}{9450} - \frac{19\,t ^6}{51975} +
  \frac{7213\,t^8}{141891750} ,
\\
  h_{11} &\approx&  \frac{1}{15}\left(1+\frac{t^2}{3}
  -\frac{t^4}{45}+\frac{2t^6}{945}-\frac{t^8}{4725}\right),\\
   h_{12} &\approx& \frac{2}{15}+\frac{t^2}{126}-\frac{19t^4}{14175}
  +\frac{181 t^6}{935550}-\frac{5443t^8}{212837625},
  \\
   h_{22}
  &\approx& \frac{1}{15}-\frac{t^2}{70}+\frac{23t^4}{9450}
  -\frac{19 t^6}{51975}+\frac{7213t^8}{141891750},
\\
  f_{11}
   &\approx&\frac{1}{15}-\frac{t^2}{105}
  +\frac{2t^4}{1575}-\frac{t^6}{6237}+\frac{1382t^8}{70945875},\\
  f_{12} &\approx& \frac{2}{15}-\frac{13t^2}{315}+\frac{253t^4}{28350}
  -\frac{ t^6}{630}-\frac{106643t^8}{425675250},
  \\ \label{k1}
  f_{22}
  &\approx& \frac{1}{15}-\frac{2t^2}{63}+\frac{2t^4}{225}
  -\frac{4 t^6}{2079}+\frac{1382t^8}{3869775}.
\end{eqnarray}
Finally, using (\ref{g})-(\ref{k1}) we easily get
\begin{eqnarray}\nonumber
\kappa_{1}^{(0)} &=& \frac{e^2}{2\pi ^2} (z_{1}+z_{2}) \left(
\frac{1}{45}\left(\frac{eB}{m^2}\right)^2 -
\frac{2}{105}\left(\frac{eB}{m^2}\right)^4
    + \frac{16}{315}\left(\frac{eB}{m^2}\right)^6
   - \frac{80}{297}\left(\frac{eB}{m^2}\right)^8+...\right)
\\
    \kappa_{2}^{(0)} &=& \frac{e^2}{2\pi ^2} \left[\frac{z_1}{6}
\left(-\frac{1}{3}\left(\frac{eB}{m^2}\right)^2
+\frac{2}{15}\left(\frac{eB}{m^2}\right)^4-\frac{16}{63}\left(\frac{eB}{m^2}\right)^6
+\frac{16}{15}\left(\frac{eB}{m^2}\right)^8+...\right)+\right.
\\ & &
\hspace{1cm} \left.+ z_{2}
 \left(\frac{1}{45}\left(\frac{eB}{m^2}\right)^2
-\frac{2}{105}\left(\frac{eB}{m^2}\right)^4+\frac{16}{315}\left(\frac{eB}{m^2}\right)^6
-\frac{80}{297}\left(\frac{eB}{m^2}\right)^8+... \right)
\right],\label{seriesinB1}
\end{eqnarray}
\begin{eqnarray}\nonumber
    \kappa_{3}^{(0)} &=& \frac{e^2}{2\pi ^2} \left[z_1
 \left(\frac{1}{45}\left(\frac{eB}{m^2}\right)^2
-\frac{2}{105}\left(\frac{eB}{m^2}\right)^4+\frac{16}{315}\left(\frac{eB}{m^2}\right)^6
-\frac{80}{297}\left(\frac{eB}{m^2}\right)^8 +...\right)+\right.
\\ & &
\hspace{1.8cm} \left.+ z_{2}
 \left(\frac{1}{15}\left(\frac{eB}{m^2}\right)^2
-\frac{2}{21}\left(\frac{eB}{m^2}\right)^4+\frac{16}{45}\left(\frac{eB}{m^2}\right)^6
-\frac{80}{33}\left(\frac{eB}{m^2}\right)^8+... \right) \right].
\label{seriesinB2}
\end{eqnarray}

and

\begin{eqnarray}\nonumber
    \kappa_{1}^{(1)} &=& \frac{e^2}{4\pi ^2 m^2}
    \left[z_{1}^2\left(\frac{1}{15}- \frac{2}{105}\left(\frac{eB}{m^2}\right)^2 +
    \frac{16}{525}\left(\frac{eB}{m^2}\right)^4- \frac{80}{693}\left(\frac{eB}{m^2}\right)^6
      +\frac{176896}{225225}\left(\frac{eB}{m^2}\right)^8+...\right)+\right.
\\ \nonumber & &
\hspace{1cm} \left.+ z_{1}z_{2}
 \left(\frac{2}{15}-
  \frac{1}{21}\left(\frac{eB}{m^2}\right)^2+ \frac{4}{45}\left(\frac{eB}{m^2}\right)^4
   - \frac{1312}{3465}\left(\frac{eB}{m^2}\right)^6+ \frac{58048}{20475}\left(\frac{eB}{m^2}\right)^8+...
   \right)
+\right.\\
& & \hspace{.1cm} \left.+ z_{2}^2
 \left( \frac{1}{15}-
  \frac{1}{35}\left(\frac{eB}{m^2}\right)^2 + \frac{92}{1575}\left(\frac{eB}{m^2}\right)^4
  - \frac{304}{1155}\left(\frac{eB}{m^2}\right)^6+ \frac{461632}{225225}\left(\frac{eB}{m^2}\right)^8+...
  \right)\right]
\end{eqnarray}
\begin{eqnarray}\nonumber
    \kappa_{2}^{(1)} &=& \frac{e^2}{4\pi ^2 m^2}
    \left[\frac{z_{1}^2}{15}\left(1+\frac{2}{3}
\left(\frac{eB}{m^2}\right)^2
-\frac{8}{15}\left(\frac{eB}{m^2}\right)^4+\frac{32}{21}\left(\frac{eB}{m^2}\right)^6
-\frac{128}{15}\left(\frac{eB}{m^2}\right)^8+...\right)+\right.
\\ \nonumber & &
\hspace{1cm} \left.+ z_{1}z_{2}
 \left(\frac{2}{15}+\frac{1}{63}\left(\frac{eB}{m^2}\right)^2
-\frac{152}{4725}\left(\frac{eB}{m^2}\right)^4+\frac{1448}{10395}\left(\frac{eB}{m^2}\right)^6
-\frac{696704}{675675}\left(\frac{eB}{m^2}\right)^8+... \right)
+\right.\\
& & \hspace{1cm} \left.+ z_{2}^2
 \left(\frac{1}{15}-\frac{1}{35}\left(\frac{eB}{m^2}\right)^2
+\frac{92}{1575}\left(\frac{eB}{m^2}\right)^4-\frac{304}{1155}\left(\frac{eB}{m^2}\right)^6
+\frac{461632}{225225}\left(\frac{eB}{m^2}\right)^8+...
\right)\right]
\end{eqnarray}
\begin{eqnarray}\nonumber
    \kappa_{3}^{(1)} &=& \frac{e^2}{4\pi ^2 m^2}
    \left[z_{1}^2\left(\frac{1}{15}-\frac{2}{105}
\left(\frac{eB}{m^2}\right)^2
+\frac{16}{525}\left(\frac{eB}{m^2}\right)^4-\frac{80}{693}\left(\frac{eB}{m^2}\right)^6
+\frac{176896}{225225}\left(\frac{eB}{m^2}\right)^8+...\right)+\right.
\\ \nonumber & &
\hspace{1cm} \left.+ z_{1}z_{2}
 \left(\frac{2}{15}-\frac{26}{315}\left(\frac{eB}{m^2}\right)^2
+\frac{1012}{4725}\left(\frac{eB}{m^2}\right)^4-\frac{8}{7}\left(\frac{eB}{m^2}\right)^6
+\frac{6825152}{675675}\left(\frac{eB}{m^2}\right)^8+... \right)
+\right.\\
& & \hspace{1cm} \left.+ z_{2}^2
 \left(\frac{1}{15}-\frac{4}{63}\left(\frac{eB}{m^2}\right)^2
+\frac{16}{75}\left(\frac{eB}{m^2}\right)^4-\frac{320}{231}\left(\frac{eB}{m^2}\right)^6
+\frac{176896}{12285}\left(\frac{eB}{m^2}\right)^8+...
\right)\right]
\end{eqnarray}
Note that in the zero field limit $B=0$, terms with the factor
$(z_1 + z_2)^2$ remain as non-zero as it is expected in quantum
electrodynamics \cite{Fradkin}

For the particular case $\textbf{k}=\textbf{k}_{\bot}$ by taking
the first terms of the expansion in powers of $B^2$ in
(\ref{seriesinB1}),(\ref{seriesinB2}) we have
\begin{eqnarray}\nonumber
 \kappa_{2}^{(0)} &=& \frac{e^4 B^2}{36\pi^2 m^4}
(\omega^2+\frac{2}{5}k_{\bot}^2 ),\\
 \kappa_{3}^{(0)} &=& \frac{e^4 B^2}{36\pi^2 m^4}
 (-\frac{2}{5}\omega^2+\frac{6}{5}k_{\bot}^2) \label{AD}
\end{eqnarray}

For the light cone $z_1 + z_2=0$, the expressions (\ref{AD}) agree
with those for the refraction indexes obtained earlier by Adler
\cite{Adler} and Dittrich \cite{Dittrich} by starting from the
Euler-Heisenberg Lagrangian and differentiating with regard to the
photon fields.

\section{Photon anomalous magnetic moment}

We differentiate with regard to $B$ the dispersion equation
$z_{1}+z_{2}=\kappa_{i}$   and get
\begin{eqnarray} \label{FAMM1} \nonumber
   \frac{ \partial z_{1}}{\partial B} &=& \frac{\partial \kappa_{i}}{\partial
   B} \\
   &=& \frac{e^2}{2\pi ^2 B}\int_{0}^{\infty}dt
\int_{-1}^{1}d\eta e^{-\frac{m^2}{eB} t} \left[ e^{\zeta
\frac{\rho_{i}}{\sinh t} } \left(\frac{m^2}{eB} t-\zeta
\right)+\frac{m^2}{eB}(z_1 +z_2)\frac{1-\eta ^2}{8} \right]
\\ \nonumber & & \hspace{1cm} +\frac{\partial z_{1}}{\partial B}\frac{e^2}{2\pi ^2
}\int_{0}^{\infty}dt \int_{-1}^{1}d\eta e^{-\frac{m^2 }{eB}t}
\left[e^{\zeta} \frac{1}{\sinh t} \left(\frac{\partial
\rho_{i}}{\partial z_{1}}-\frac{ \rho_{i}}{eB}\frac{1-\eta ^2}{4 }t
\right)+\frac{1-\eta ^2}{8 t} \right].
\end{eqnarray}

and using $\frac{\partial z_{1}}{\partial B}=-2\omega \frac{\partial
\omega}{\partial B}$ , we obtain an expression for the photon
anomalous magnetic moment (see Fig. 1)
\begin{eqnarray}\label{FAMM2}
    \mu_{\gamma}&=&-\frac{\partial \omega}{\partial B} \\
    &=& \frac{1}{2 \omega B}\frac{\frac{e^2}{2\pi ^2
    B}\int_{0}^{\infty}dt
\int_{-1}^{1}d\eta e^{-\frac{m^2}{eB} t} \left[ e^{\zeta
\frac{\rho_{i}}{\sinh t} } \left(\frac{m^2}{eB} t-\zeta
\right)+\frac{m^2}{eB}(z_1 +z_2)\frac{1-\eta ^2}{8}
\right]}{1-\frac{e^2}{2\pi ^2 }\int_{0}^{\infty}dt
\int_{-1}^{1}d\eta e^{-\frac{m^2 }{eB}t} \left[e^{\zeta}
\frac{1}{\sinh t} \left(\frac{\partial \rho_{i}}{\partial
z_{1}}-\frac{ \rho_{i}}{eB}\frac{1-\eta ^2}{4 }t
\right)+\frac{1-\eta ^2}{8 t} \right]}.
\end{eqnarray}
\begin{figure}[!htbp]
\includegraphics[width=4in]{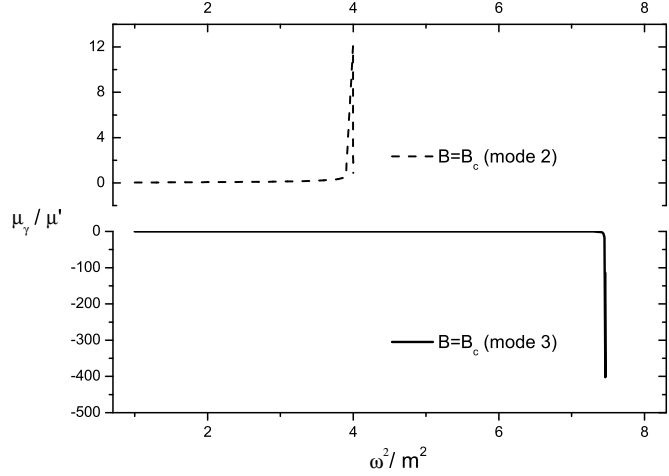}
\caption{\label{fig:Phvk} Photon magnetic moment curve drawn with
regard to squared frequency,  for the second and  the third modes
showing a peak near the first and the second threshold,
respectively.}
\end{figure}

\subsection{Small frequencies limit $z_1 ,z_2<<m^2$}
For small frequencies $m^2>>z_1 ,z_2$  we can take only the first
term $\kappa_{i}^{(0)}$ in the series expansion
(\ref{series-op-pol}). In this section we are interested in
computing $\mu_{\gamma}^{2,3}$ (perpendicular propagation).
 The solutions of the dispersion equations in this limit
$z_1+z_2=\kappa_{2,3}^{(0)}$ for  the second and the third modes,
respectively, are
\begin{eqnarray} \label{dispeq2}
 z_1 &=& -z_2\frac{1- \frac{e^2}{2\pi ^2}  \int_{0}^{\infty}dt
 e^{-\frac{m^2}{eB}t}
 g_{22}}{1-\frac{e^2}{2\pi ^2} \int_{0}^{\infty}dt
 e^{-\frac{m^2}{eB}t}
g_{21}} ,\\ \label{dispeq3}
 z_1 &=& -z_2 \frac{1-\frac{e^2}{2\pi ^2}  \int_{0}^{\infty}dt
e^{-\frac{m^2}{eB}t }g_{32}}{1-\frac{e^2}{2\pi ^2}
\int_{0}^{\infty}dt
 e^{-\frac{m^2}{eB}t}
 g_{31}}
\end{eqnarray}
 From (\ref{dispeq2}), (\ref{dispeq3}), after a straightforward calculation we get
$\mu_{\gamma}^{2,3}=-\frac{\partial \omega}{\partial
B}=\frac{1}{2\omega}\frac{\partial \kappa_{i}^{(0)}}{\partial B}$

\begin{eqnarray}
\mu_{\gamma}^{2} &=& z_2 \frac{e^2B_c}{4\pi ^2 \omega B}
\frac{\left( 1-\frac{e^2}{2\pi ^2} \int_{0}^{\infty}dt
 e^{-\frac{m^2}{eB}\tau }
g_{21}\right)  \int_{0}^{\infty}dt
 e^{-\frac{m^2}{eB}t}
 g_{22}t-\left(1- \frac{e^2}{2\pi ^2}  \int_{0}^{\infty}dt
 e^{-\frac{m^2}{eB}t}
 g_{22}\right)\int_{0}^{\infty}dt
 e^{-\frac{m^2}{eB}t }
g_{21}t}{\left( 1-\frac{e^2}{2\pi ^2} \int_{0}^{\infty}dt
 e^{-\frac{m^2}{eB}t }
g_{21} \right)^2},
\\
   \mu_{\gamma}^{3} &=& z_2 \frac{e^2B_c}{4\pi ^2 \omega B}  \frac{\left(
1-\frac{e^2}{2\pi ^2} \int_{0}^{\infty}dt
 e^{-\frac{m^2}{eB}t }
g_{31} \right)  \int_{0}^{\infty}dt
 e^{-\frac{m^2}{eB}t}
 g_{32}t-\left(1- \frac{e^2}{2\pi ^2}  \int_{0}^{\infty}dt
 e^{-\frac{m^2}{eB}t}
 g_{32}\right)\int_{0}^{\infty}dt
 e^{-\frac{m^2}{eB}t }
g_{31}t}{\left( 1-\frac{e^2}{2\pi ^2} \int_{0}^{\infty}dt
 e^{-\frac{m^2}{eB}t }
g_{31} \right)^2}.
\end{eqnarray}
It is easy to see that $\mu_{\gamma}^{2,3}>0$ due to the following
relations:
\begin{eqnarray*}
  & & \frac{e^2}{2\pi ^2}<<1,
\int_{0}^{\infty}dt
 e^{-\frac{m^2}{eB}t}
 g_{21}t<0,\int_{0}^{\infty}dt
 e^{-\frac{m^2}{eB}t}
 g_{22}t,\int_{0}^{\infty}dt
 e^{-\frac{m^2}{eB}t}
 g_{31}t,\int_{0}^{\infty}dt
 e^{-\frac{m^2}{eB}t}
 g_{32}t>0,\\
& & \left|\int_{0}^{\infty}dt
 e^{-\frac{m^2}{eB}t}
 g_{21}t\right|>\left|\int_{0}^{\infty}dt
 e^{-\frac{m^2}{eB}t}
 g_{32}t\right|>\left|\int_{0}^{\infty}dt
 e^{-\frac{m^2}{eB}t}
 g_{22}t\right|=\left|\int_{0}^{\infty}dt
 e^{-\frac{m^2}{eB}t}
 g_{31}t\right|
 \end{eqnarray*}
 \section{Acknowledgements}
 The authors thank A.E.Shabad for useful comments and S.Villalba
 for a discussion. They thank also OEA-ICTP for support under Net-35.

\end{document}